# Intrinsic defects and conduction characteristics of $Sc_2O_3$ in thermionic cathode systems

Ryan M. Jacobs,[*] John H. Booske,[*,†] and Dane Morgan[*,‡]
[*]Interdisciplinary Materials Science Program, University of Wisconsin-Madison, 1509 University Avenue, Madison, Wisconsin 53706, USA
[†]Department of Electrical and Computer Engineering, University of Wisconsin-Madison, 1415 Engineering Drive, Madison, Wisconsin 53706, USA
[‡]Department of Materials Science and Engineering, University of Wisconsin-Madison, 1509 University Avenue, Madison, Wisconsin 53706, USA

Recent experimental observations indicate that bulk $Sc_2O_3$ (~200 nm thick), an insulator at room temperature and pressure, must act as a good electronic conductor during thermionic cathode operation, leading to the observed high emitted current densities and overall superior emission properties over conventional thermionic emitters which do not contain $Sc_2O_3$. Here, we employ *ab initio* methods using both semilocal and hybrid functionals to calculate the intrinsic defect energetics of Sc and O vacancies and interstitials and their effects on the electronic properties of $Sc_2O_3$ in an effort to explain the good conduction of $Sc_2O_3$ observed in experiment. The defect energetics were used in an equilibrium defect model to calculate the concentrations of defects and their compensating electron and hole concentrations at equilibrium. Overall, our results indicate that the conductivity of $Sc_2O_3$ solely due to the presence of intrinsic defects in the cathode operating environment is unlikely to be high enough to maintain the magnitude of emitted current densities obtained from experiment, and that presence of impurities are necessary to raise the conductivity of $Sc_2O_3$ to a high enough value to explain the current densities observed in experiment. The necessary minimum impurity concentration to impart sufficient electronic conduction is very small (approximately $7.5 \times 10^{-3}$ ppm) and is probably present in all experiments.

## I. INTRODUCTION

Electron emitter cathodes have myriad applications in military and civilian infrastructure, industrial manufacturing and advanced scientific research. They are critical components in radar, communications, broadcast media, medical imaging, materials characterization, materials processing, domestic, commercial and industrial food production, electronic warfare, and high energy physics research technologies.[1]

Scandate ($Sc_2O_3$-containing) cathodes have attracted significant attention lately in both academic and industrial research settings due to their generally superior properties over other thermionic electron emitters. These superior characteristics include: higher electron emission, lower operating temperature, and good resistance to surface contamination.[2] Lower operating temperatures promote less degradation of the cathode (e.g. depletion of surface Ba by evaporation at high temperatures) which could in turn lead to longer cathode lifetimes while simultaneously generating a higher density of emitted electrons. A large supply of emitted electrons is important in applications demanding high current and high power densities such as THz-regime vacuum electronic devices (VED's).[1,3]

Common thermionic emitters such as the dispenser B-type cathode are composed of pressed and sintered polycrystalline W powder impregnated with a precise compositional mixture of BaO, CaO and $Al_2O_3$, with typical molar ratios of 4-1-1 or 5-3-2. Scandate cathodes augment this composition by also including a small amount of $Sc_2O_3$ to the emissive mix.[4] One drawback to scandate cathodes prepared in this fashion is an observed non-uniformity in electron emission. However, recent alterations to the cathode processing techniques appear to have helped mitigate this issue.[5,6] In addition to the pressed and sintered dispenser-type scandate cathode, thin film and top layer scandate cathodes have also been studied. One particularly interesting recent investigation has used a polycrystalline W foil as a base with layers of sputter-deposited $Sc_2O_3$ and BaO on top of the W foil.[7] Top layer cathodes generally consist of a deposited $Sc_2O_3$ layer on the sintered and impregnated W powder substrate.[8,9]

In addition to thermionic emission cathodes, $Sc_2O_3$ has numerous other applications. $Sc_2O_3$ has been used as a thin film gate oxide on GaN for high electron mobility transistors due to its high dielectric constant and better lattice matching over other gate insulators such as $Gd_2O_3$, $SiO_2$, and $Si_3N_4$. Because of the improved lattice matching, the use of $Sc_2O_3$ as the gate oxide layer has improved device performance by minimizing current leakage and interface state density.[10-12] Due to its high refractive index, $Sc_2O_3$ has been shown to be a superior antireflective coating for UV laser[13,14] and superluminescent LED applications.[15] Overall, $Sc_2O_3$ is versatile and has attracted a great amount of interest in numerous advanced materials research fields.

This particular investigation is motivated by the recent results of experimental studies of electron emission from scandia-coated cathodes reported in Ref. 7. In Ref. 7, $Sc_2O_3$ was sputter deposited onto a polycrystalline W foil. BaO was subsequently deposited on top of some of the $Sc_2O_3$ regions in a square pattern with a final film thickness on the order of a couple hundred nanometers. The emission characteristics of the samples were studied with photoelectron emission microscopy (PEEM) and thermionic emission microscopy (ThEEM). Overall, the cathode areas which contained film layers of W/$Sc_2O_3$/BaO exhibited much brighter and highly localized emission than the areas containing only W/BaO layers. The observed higher intensity corresponds to higher electron emission from the $Sc_2O_3$-containing areas. These results are interesting namely because one would expect $Sc_2O_3$ to be insulating due to it having a large bandgap of roughly 6 eV.[16-18] $Sc_2O_3$ films that are greater than 100 nm in thickness should be expected to behave as bulk-like $Sc_2O_3$.

There have been numerous attempts to describe the physics behind the enhanced emission of scandate cathode systems. These theories include the formation of Sc-O dipole layers on the W surface which lower the work function,[19,20] a semiconductor model whereby an applied potential lowers the emission barrier near the $Sc_2O_3$ surface,[21] and Ba-Sc-O surface complexes that also act to reduce the work function.[22-24] Although numerous theories currently exist, it is important to note that an unequivocal fundamental physical understanding of enhanced emission from $Sc_2O_3$ is still missing.

The overall purpose of this paper is to investigate mechanisms by which bulk-like $Sc_2O_3$ (an insulator) could act as a good conductor in the cathode environment ($T \sim 1200$ K, $P \sim 10^{-10}$ Torr). We first consider intrinsic defects by using *ab initio* methods to calculate the defect thermodynamics and effects these defects have on the electronic properties of $Sc_2O_3$. We then briefly consider the role of impurities. This serves as the first step in understanding the enhanced emission for $Sc_2O_3$.

## II. COMPUTATIONAL METHODS AND THEORETICAL BACKGROUND

### A. DFT Methods and $Sc_2O_3$ structure

All calculations of total supercell energies and defect energetics were performed using Density Functional Theory as implemented by the Vienna *ab initio* simulation package (VASP)[25] with a plane wave basis set. The electron exchange and correlation functionals were treated with the generalized gradient approximation (GGA)[26] and projector augmented wave (PAW)- type pseudopotentials[27] for both Sc and O atoms. The valence electron configuration of the Sc and O atoms utilized in all calculations were $3p^6 4s^2 3d^1$ and $2s^2 2p^4$, respectively. The plane wave cutoff energy was set to 290 eV and reciprocal space integration in the irreducible Brillouin Zone was conducted with the Monkhorst-Pack scheme.[28] A 4x4x4 k-point mesh was used for the 1x1x1 (40 atoms) primitive cell of $Sc_2O_3$. Due to the high computational expense of larger supercells, a smaller k-point mesh of 1x1x1 k-points was used for the 2x2x2 (320 atoms) $Sc_2O_3$ supercell. It was verified that at higher k-point densities, the total $Sc_2O_3$ supercell energy changed less than 1 meV/f.u. (f.u. = formula unit), indicating that the chosen k-point densities were sufficient to ensure accurate total energies. The accuracy of the 4x4x4 k-point mesh is consistent with a previous *ab initio* study on $Sc_2O_3$.[29] In addition to the use of GGA, the hybrid HSE functional was employed with 25% Hartree-Fock exchange.[30] Since the band gap underestimation prevalent in LDA and GGA can potentially affect the defect energetics and concentrations,[31] the use of HSE will provide the most rigorous quantitative results. GGA is also implemented to obtain fast, qualitative results that still prove useful for comparison with both our calculated HSE values and GGA defect energetics of other materials in the literature.

$Sc_2O_3$ crystallizes in the cubic bixbyite crystal structure (space group 206, $Ia\bar{3}$). In a previous *ab initio* study[29] it was found that other polymorphs of $Sc_2O_3$ become stable under conditions of high pressure, high temperature, or a combination of both. However, the cubic bixbyite phase is thermodynamically stable at standard temperature and pressure conditions and also under typical scandate cathode operating conditions of high temperature (T~1200 K) and low pressure (P~$10^{-10}$ Torr).[29,32] Therefore, our conditions of interest rule out all $Sc_2O_3$ polymorphs except the cubic bixbyite phase. Fig. 1 depicts the conventional cubic bixbyite structure of $Sc_2O_3$. The primitive cell contains 40 atoms, while this conventional unit cell has 80 atoms. Sc resides on the 8b and 24d Wyckoff sites while O occupies the 48e sites. The 8b and 24d sites are both tetrahedrally coordinated by four O atoms, however the 8b sites have higher overall symmetry. The 48e sites are octahedrally coordinated by six Sc atoms. In addition, there are natively vacant anion sites at the 8a and 16c Wyckoff positions. Filling the 16c sites of the conventional unit cell with O atoms would yield a 2x2x2 fluorite structure.[33]

The *ab initio* fully relaxed lattice parameters are a = b = c = 9.871(65) Å and $\alpha = \beta = \gamma = 90$ degrees. These results are consistent with x-ray diffraction data[34] and previous *ab initio* studies on this material.[18,29] The electronic band gap as predicted by GGA is about 3.9 eV, consistent with Ref. 29 but significantly below the experimentally determined band gap of 5.7-6 eV. However, the electronic band gap predicted by the HSE hybrid functional is 5.7 eV, within the experimentally determined range. Finally, the formation energy of $Sc_2O_3$ from pure Sc metal and $O_2$ gas (evaluated at T = 298K and $P_{O2}$ = 1atm) was calculated to be -1861 kJ/mol for GGA and -1938 kJ/mol for HSE, in reasonable agreement with the experimental result of -1908.8 kJ/mol (also for 298K and $P_{O2}$ = 1atm).[35]

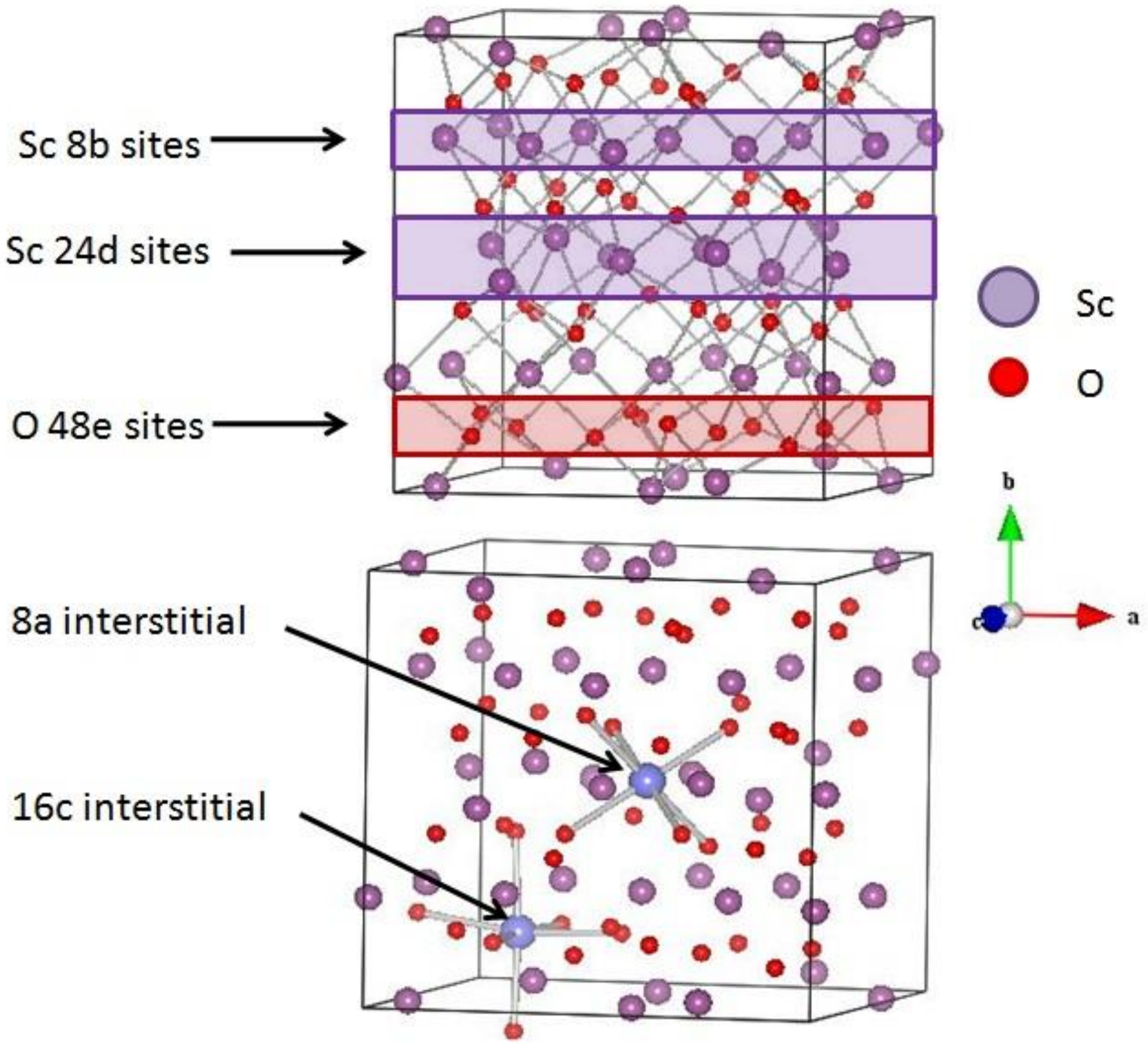

FIG. 1. $Sc_2O_3$ cubic bixbyite conventional unit cell depicting native and interstitial Wyckoff positions used in this study.

### B. Range of Chemical Potentials

For the purposes of calculating the defect energetics and corresponding defect concentrations, it is necessary to examine the relevant range of chemical potentials for Sc and O which comprise the $Sc_2O_3$ system. The values of the Sc and O chemical potentials are needed to perform the defect formation energy calculations. We accomplish this by using a similar formalism as developed in previous studies.[36-40] The chemical potentials of Sc and O depend on one another through the cohesive energy of $Sc_2O_3$ $E^{fu}_{Sc_2O_3}$, which is expressed (per formula unit) as:

$$2\mu_{Sc}^{Sc_2O_3} + 3\mu_O^{Sc_2O_3} = E^{fu}_{Sc_2O_3}, \qquad (1)$$

and can be directly obtained from VASP calculations. In Eq. (1), $\mu_O^{Sc_2O_3}$ is the chemical potential of O in $Sc_2O_3$ and $\mu_{Sc}^{Sc_2O_3}$ is the chemical potential of Sc in $Sc_2O_3$. In the cathode operating environment of ~1200 K and $10^{-10}$ Torr, it is considered that Sc originates from its stable oxide (i.e. $Sc_2O_3$) and O originates from a reservoir of $O_2$ gas. This assumption is consistent with our calculations of Sc metal, which predict that Sc metal is not thermodynamically stable at 1200 K and $10^{-10}$ Torr (a similar Sc source was used in Ref. 22). For $Sc_2O_3$ to be stable with respect to Sc metal, the following inequality must hold:

$$\mu_{Sc}^{Sc_2O_3} \leq \mu_{Sc}^0, \qquad (2)$$

where $\mu_{Sc}^{0}$ is the chemical potential of Sc referenced to pure Sc metal. The other necessary condition for stable $Sc_2O_3$ is that it be stable against the loss of O:

$$\mu_O^{Sc_2O_3} \le \mu_O^0 , \qquad (3)$$

where $\mu_O^0$ is the chemical potential of O referenced to pure $O_2$ gas. Using Eqs. (1-3), the formation energy for $Sc_2O_3$ from Sc metal and $O_2$ gas can be expressed in terms of the above chemical potentials:

$$\Delta E_{form}^{Sc_2O_3} = 2\mu_{Sc}^{Sc_2O_3} + 3\mu_O^{Sc_2O_3} - 2\mu_{Sc}^0 - 3\mu_O^0 , \quad (4)$$

Using Eq. (1) and equilibrium of O between $O_2$ gas and $Sc_2O_3$, one can express the relevant Sc chemical potential as a function of the $Sc_2O_3$ cohesive energy (per formula unit) and O chemical potential:

$$\mu_{Sc}^{Sc_2O_3} = \frac{1}{2}\left(E_{Sc_2O_3}^{fu} - 3\mu_O^0\right). \qquad (5)$$

The O chemical potential can be calculated by using a combination of VASP total energies and experimental thermodynamic data for $O_2$ gas at the relevant reference state.[41] It takes the form:[42]

$$\mu_O^0 = \frac{1}{2}\left[E_{O_2}^{VASP} + \Delta h_{O_2}^0 + \left(H\left(T,P^0\right) - H\left(T^0,P^0\right) - TS\left(T,P^0\right)\right) + kT\ln\left(\frac{P}{P^0}\right) - \left(G_{O_2}^{s,vib}(T) - H_{O_2}^{s,vib}(T^0)\right)\right], \quad (6)$$

where $E_{O2}^{VASP}$ is the *ab initio* calculated energy of an $O_2$ gas molecule, $\Delta h_{O2}^0$ is a numerical correction for the overbinding of the $O_2$ molecule in DFT, $H(T,P^0)$ and $H(T^0,P^0)$ are the gas enthalpy values at standard and general temperatures $T^0$ and $T$, respectively, $S(T,P^0)$ is the gas entropy, and the logarithmic term is the adjustment of the chemical potential for arbitrary pressure. The final terms in Eq. (6), $G_{O2,s,vib}$ and $H_{O2,s,vib}$, shift the value of $\mu_O^0$ to account for solid phase vibrations, which are approximated with an Einstein model with an Einstein temperature of 500 K following Refs. 24 and 42. The value of $\Delta h_{O2}^0$ takes into account the temperature increase of $O_2$ gas from 0 K to $T^0$, the contribution to the enthalpy at $T^0$ when oxygen is in the solid phase, and also corrects for the numerical error in DFT from overbinding the $O_2$ molecule. Because $\Delta h_{O2}^0$ is obtained from comparing calculated formation energies and experimental formation enthalpies of numerous oxides, it is dependent on the pseudopotential and exchange/correlation functional used in the calculations. For our GGA defect energetics, we use the value of $\Delta h_{O2}^0$ = 0.33 eV/$O_2$ from Ref. 42, however no such correction has been formulated for HSE.

Fig. 2 compares the calculated formation energies of several non-transition metal oxides (with both GGA and HSE) with their respective experimental formation enthalpies. For our GGA values we obtain a value of $\Delta h_{O2}^0$ = 0.35 eV/$O_2$. The difference of 20 meV from the result of 0.33 eV/$O_2$ in Ref. 42 is due to the use of a different set of oxide materials in the calculations. For our HSE values we obtain $\Delta h_{O2}^0$ = 0.33 eV/$O_2$. Interestingly, this shows that the DFT errors for modeling the $O_2$ molecule are related more to the pseudopotential used and not the exchange/correlation functional.

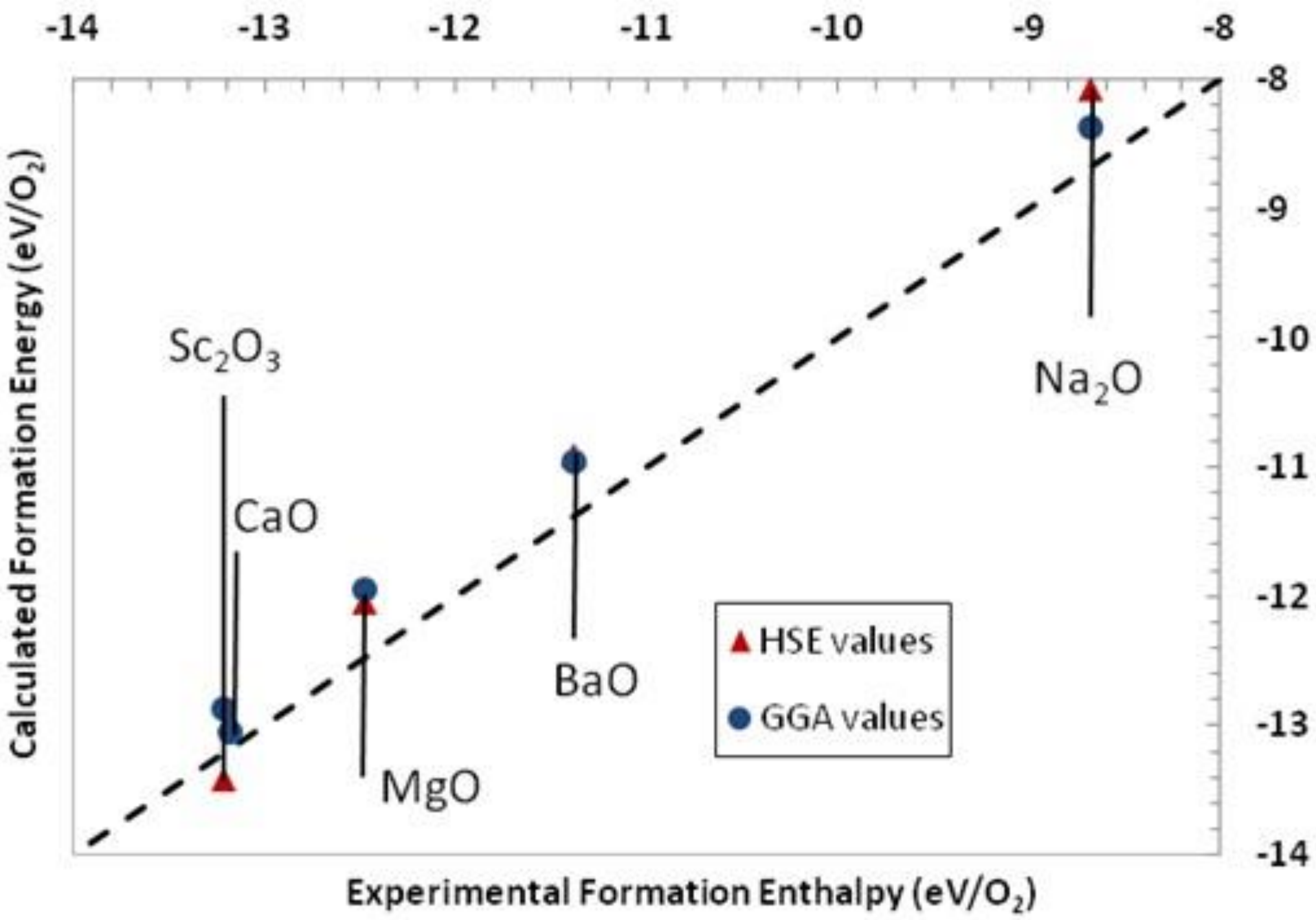

FIG. 2. Comparison of calculated formation energies to experimental formation enthalpies under standard conditions for several oxides. The average energy difference between the calculated and experimental values yields the value of $\Delta h^0_{O2}$.

Combining Eqs. (2-4) yields the bound on the Sc chemical potential consistent with stable $Sc_2O_3$ and from which the defect energetics will be evaluated:

$$\mu^0_{Sc} + \frac{1}{2}\Delta E^{Sc_2O_3}_{form} \leq \mu^{Sc_2O_3}_{Sc} \leq \mu^0_{Sc}\,, \qquad (7)$$

This physical bound on the Sc (and thus O) chemical potentials provide an overall stability limit for $Sc_2O_3$ in our system. We choose three conditions within this bound to evaluate the defect energetics of the system: $\mu^{Sc_2O_3}_{Sc} = \mu^0_{Sc}$ (Sc rich conditions, high Sc partial pressure and low O partial pressure), $\mu^{Sc_2O_3}_{O} = \mu^0_{O}$ (O rich conditions, high O partial pressure and low Sc partial pressure), intermediate $\mu^{Sc_2O_3}_{Sc}$ and $\mu^{Sc_2O_3}_{O}$ appropriate for cathode operating conditions (T= 1200 K and P = $10^{-10}$ Torr). Calculation of these chemical potentials is summarized in Table I.

### C. Defects in $Sc_2O_3$

The concentration of a crystal defect of type $\alpha$ at equilibrium and in the dilute limit takes the form

$$C_\alpha = N_\alpha exp\left(\frac{-\Delta G^\alpha_{form}}{k_B T}\right), \qquad (8)$$

Table I. External Sc and O chemical potentials under different conditions within the relevant chemical potential range. The values for intermediate $\mu^{Sc_2O_3}_{Sc}$ and $\mu^{Sc_2O_3}_{O}$ correspond to typical scandate cathode operating conditions of T=1200 K and P=$10^{-10}$ Torr. The GGA values are listed first and the HSE values second.

| Conditions | $\mu_{Sc}^{Sc_2O_3}$ (eV/Sc) | $\mu_{O}^{Sc_2O_3}$ (eV/O) |
|---|---|---|
| O rich (Sc poor) | $\frac{1}{2}(E_{Sc_2O_3}^{f.u.} - 3\mu_O^0) =$ -16.675/ -18.284 | $\mu_O^0$ = -4.016/ -6.957 |
| Cathode operating conditions | $\frac{1}{2}(E_{Sc_2O_3}^{f.u.} - 3\mu_O^0)$ (1200 K,$10^{-10}$ Torr)) = -13.871/ -15.480 | $\mu_O^0$ (1200 K,$10^{-10}$ Torr) = -5.886/ -8.827 |
| Sc rich (O poor) | $\mu_{Sc}^0$ = -6.679/ -8.228 | $\frac{1}{3}(E_{Sc_2O_3}^{f.u.} - 2\mu_{Sc}^0)$ = -10.620 / -13.661 |

where $\Delta G_{form}^{\alpha}$ is the free energy of forming the defect, $N_\alpha$ is the number of sites per unit volume the defect can occupy, $k_B$ is the Boltzmann constant and $T$ is the absolute temperature. The free energy of formation can be expanded into

$$\Delta G_{form}^{\alpha} = \Delta E_{form}^{\alpha} - T\Delta S_{form}^{\alpha} + P\Delta V_{form}^{\alpha}, \qquad (9)$$

$\Delta E_{form}^{\alpha}$ is the change in total energy of the system by introduction of the defect (see Eq. (10)), $\Delta S_{form}^{\alpha}$ is the formation entropy with configurational terms, which we assume is primarily due to vibrational effects, $\Delta V_{form}^{\alpha}$ is the volume change when a defect is introduced, and for relevant pressures its contribution to $\Delta G_{form}^{\alpha}$ can be ignored. These assumptions lead us to consider the defect formation energies as consisting of total energy differences, including the external chemical potentials and Fermi energy adjustments if the defect is charged.

In order to determine the possible role defects play in the superior emission of scandia-coated cathodes as reported in Ref. 7, the defect energetics were calculated using *ab initio* methods for the three conditions developed in Table I, and the relevant defect concentrations were calculated. Neutral and charged vacancies on the 8b and 24d Sc sites and the 48e O site were investigated. The relevant interstitial sites are at the 8a and 16c positions. Both neutral and charged Sc and O interstitials were investigated. Numerous charge states were considered in order to find the most stable charged defect species. In the case of O interstitials, charge states of 0, 1-, 2- and 3- were investigated, while charges of 0, 1+, 2+ and 3+ were used for the Sc interstitials. Sc vacancies were investigated with charge states of 0, 1-, 2- and 3- while O vacancies had charge states of 0, 1+ and 2+. Higher charge states were briefly investigated with only GGA for the 3+ O vacancy, 4+ Sc interstitial, 4- Sc vacancy and 4- O interstitial. However, these higher charge states are not stable defect charge states. In the case of the 3+ O vacancy and 4+ Sc interstitial, the defect formation energy is higher than the 2+ O vacancy and 3+ Sc interstitial under p-type conditions, indicating that these defects are not stable. In the case of the 4- Sc vacancy and 4- O interstitial, the extra charge is no longer associated with the defect, rather it is delocalized in the conduction band of the crystal, demonstrating that one cannot charge the defect to this level. Lastly, we simulated a 1+ O interstitial to examine whether a defect that typically has a negative charge could instead have a positive charge. In this case the formation energy is larger than even the neutral O interstitial, indicating that this defect type is also unstable. Defect calculations used an isolated defect method in which only one defect was introduced per supercell, and no defect complexes or clusters were considered. Supercell sizes of 1x1x1 and 2x2x2 were used in scaling calculations to extrapolate the defect formation energies to infinite supercell size (see Eq. (12)). For the singly-charged defects, electrons were doped into the supercell if the defect was negatively charged (e.g. O interstitial) and electrons were removed from the supercell if the defect was positively charged (e.g. O vacancy). The defect formation energy $\Delta E_{form}^{\alpha}$ of a general defect of type α and charge $q$ is:

$$\Delta E^{\alpha}_{form} = E^{\alpha}_{def} - E_{perf} \pm \mu_{\alpha} + q\mu_{e}, \quad (10)$$

where $E^{\alpha}_{def}$ is the energy of the defected supercell, $E_{perf}$ is the energy of the undefected supercell, $\mu_{\alpha}$ is the chemical potential of the defect species (+ sign for vacancies, - sign for interstitials), and $q\mu_e$ is the electron chemical potential multiplied by the charge state of the defect. The electron chemical potential can be further developed as:

$$q\mu_e = q\left(E_{VBM} + E_{Fermi} + E_{shift}\right), \quad (11)$$

where $E_{VBM}$ is the energy of the valence band maximum, $E_{Fermi}$ is the Fermi energy referenced to the bulk undefected supercell, and $E_{shift}$ is the electrostatic potential energy shift. For convenience, the Fermi level is referenced to the valence band maximum in all defect calculations. The electrostatic energy shift serves as a correction for the shift of the valence band maximum by introduction of charged defect species in a finite sized supercell. Different values of $E_{shift}$ can be applied from the VASP calculations, and for our calculations we use a modified version of the method employed by Lany and Zunger in Ref. 43. In their work, the electrostatic energy shift was obtained by calculating the average difference between the atomic-site electrostatic potentials of the defected and perfect supercells, excluding the immediate neighbors of the defect. Our method uses this averaging technique, however we also include the potentials of the nearest neighbor atoms to the defect, as this method was found to be the most accurate correction for our system.[44] Overall, the energy scale of the electrostatic energy shift is small, on the order of $10^{-3}$ to $10^{-1}$ eV, so the exact method of treating this shift will not qualitatively impact our results.

In addition to this correction for the shifting of the valence band maximum from that of the perfect supercell, the defect formation energies must also be corrected for the finite size effects of the supercell approach. Although the isolated defect method reduces defect-defect interactions within the supercell, there are still spurious strain and electrostatic interactions between the periodic supercell images. The charge corrections, first developed by Leslie and Gillan[45] and further by Makov and Payne[46] (which scale as $1/L$ and $1/L^3$) and the strain corrections (which also scale as $1/L^3$)[47] can be simplified to the form in Refs. 44,48,49. Thus, the general dependence of defect formation energy with supercell size can be expressed as:

$$\Delta E^{\alpha}_{form,sc} = \Delta E^{\alpha}_{form,infinite} + a\left(\frac{1}{L}\right) + b\left(\frac{1}{L^3}\right), \quad (12)$$

where $\Delta E^{\alpha}_{form,sc}$ is the defect formation energy in the supercell, $\Delta E^{\alpha}_{form,infinite}$ is the defect formation energy of a hypothetical infinite crystal, $L$ is the supercell size, and $a$ and $b$ are fitting constants. The inverse cubic term is sufficiently small that we ignore it in our finite size scaling. The constants $a$ and $b$ are sometimes represented as analytic electrostatic expressions which incorporate the Madelung constant of the supercell, the dielectric constant of the material and the defect charge. However, these purely electrostatic expressions have been shown to be unreliable for numerous systems, as elastic effects are ignored and the effects of defect charge are oversimplified.[50-52] We thus follow the methods used in Refs. 44,48,52 to obtain more realistic defect formation energies in the dilute limit. By running both the 1x1x1 and 2x2x2 supercells, two defect formation energies are obtained (effectively for two different defect concentrations, since the supercell size is changing). Therefore, these two formation energies define a linear plot of $\Delta E^{\alpha}_{form,sc}$ versus $1/L$ where the constant $a$ is the slope and $\Delta E^{\alpha}_{form,infinite}$ is the y-intercept. The value of $\Delta E^{\alpha}_{form,infinite}$ is the defect formation energy in the dilute limit. We assume that the $b/L^3$ term makes a minimal contribution as the larger calculation cells needed for fitting this term are not computationally practical at this time.

## III. RESULTS AND DISCUSSION

Table II contains the GGA and HSE defect formation energies for every defect calculated under the different external conditions listed in Table I. Fig. 3 consists of the data from Table II plotted as a function of the Fermi energy for the case of intermediate $\mu_{Sc}^{Sc_2O_3}$ and $\mu_{O}^{Sc_2O_3}$ appropriate for cathode operating conditions. The defect formation energy is a function of the Fermi energy and varies linearly with the charge state of the defect in question. Plots such as those shown in Fig. 3 are very useful when analyzing intrinsic defects because it allows one to easily see which defect species are the most stable based on the position of the Fermi energy in equilibrium. The position of the Fermi energy can be dictated by several things, such as external dopants, electrical contact with another material, or simply from the presence of intrinsic crystal defects. For our study we focus on calculating the position of the Fermi energy from the presence of just intrinsic defects in $Sc_2O_3$ under equilibrium conditions. From inspection of the plots in Fig. 3 one can predict that, to a first approximation, the equilibrium Fermi energy should lie in close proximity to a crossover point where the lowest energy defect types of opposite charge states have the same formation energy. This can be reasoned based solely on the fact that the entire system must remain charge neutral.

To calculate the equilibrium defect concentrations and corresponding electronic properties which arise due to the presence of these defects, we employ similar methods as detailed in Refs. 53,54 to model the defect concentration equations and Ref. 55 for the electron and hole concentrations. The relationships between the electron concentration in the conduction band and the hole concentration in the valence band and the Fermi energy are given by:

$$[e^-] = n_c(T) = N_c(T)\,exp\left(\frac{-(E_{CBM} - E_{Fermi})}{k_B T}\right), \quad (13)$$

$$[h^+] = p_v(T) = P_v(T)\,exp\left(\frac{-(E_{Fermi} - E_{VBM})}{k_B T}\right), \quad (14)$$

where $E_{CBM}$ is the energy at the conduction band minimum (taken as the calculated band gap, $E_{gap}$= 3.9 or 5.7 eV for GGA and HSE, respectively), $E_{VBM}$ is the energy at the valence band maximum (taken as equal to 0 eV), and $N_c(T)$ and $P_v(T)$ are the effective densities of states in the conduction and valence bands, respectively, and were calculated using density of states data for a perfect 1x1x1 $Sc_2O_3$ supercell. The method of determining these effective densities of states can be found in Ref. 55 and won't be repeated here. Summing the concentrations from acceptor-type defects (negative charge) and donor-type defects (positive charge) along with their compensating electrons and holes gives the condition of charge neutrality for the system:

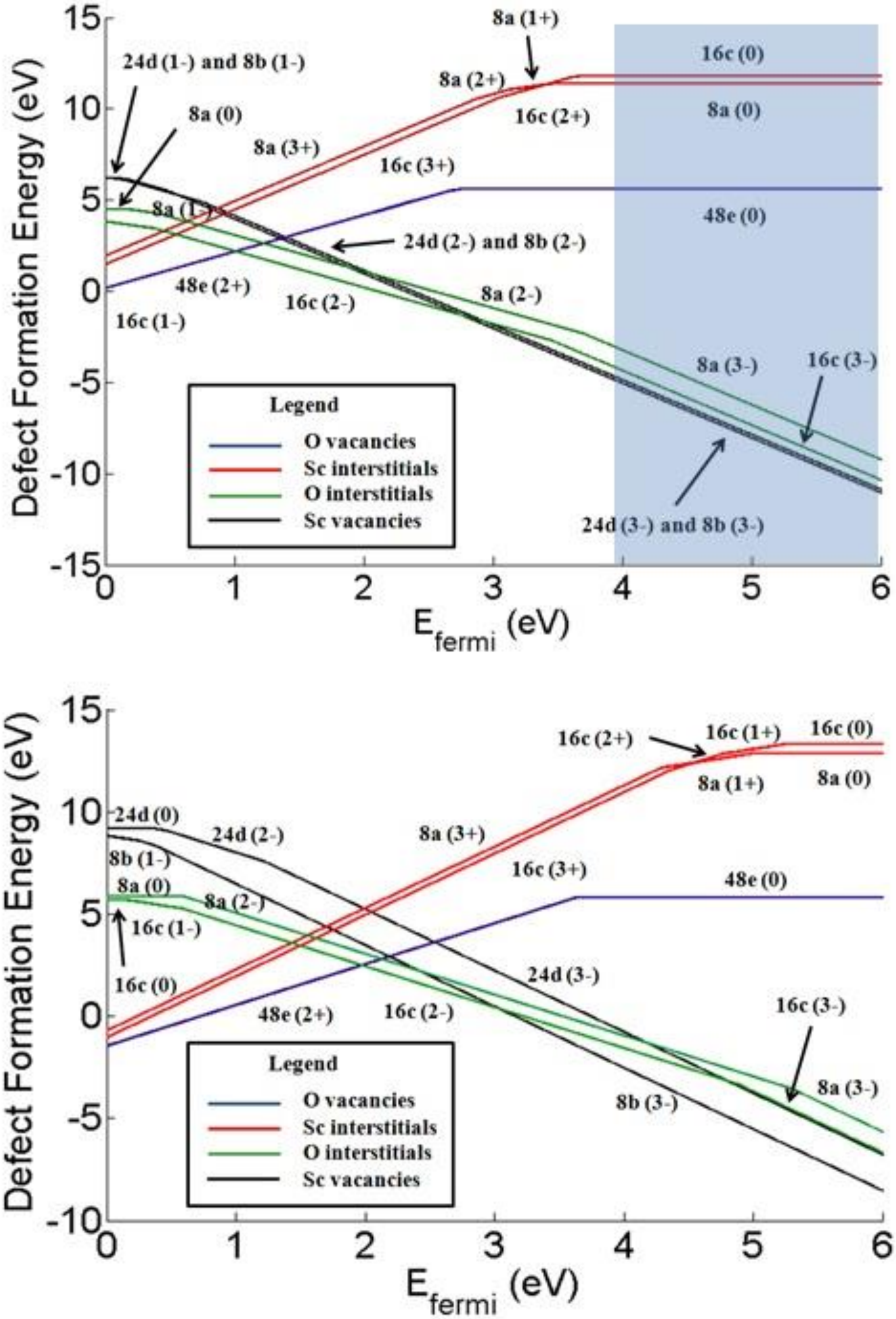

FIG. 3. Calculated defect formation energies for intrinsic defects in $Sc_2O_3$ as a function of Fermi level for (top) GGA and (bottom) HSE functionals. Intermediate values of $\mu_{Sc}^{Sc_2O_3}$ and $\mu_{O}^{Sc_2O_3}$ (cathode operating conditions: T= 1200 K and P = $10^{-10}$ Torr) were used in both cases. The shaded portion indicates Fermi energies that lie above the calculated GGA bandgap.

$$2[V_{O48e}^{2+}]+3[Sc_{i,16c}^{3+}]+3[Sc_{i,8a}^{3+}]+[h^{+}]=2[O_{i,16c}^{2-}]+2[O_{i,8a}^{2-}]+3[V_{Sc8b}^{3-}]+3[V_{Sc24d}^{3-}]+[e^{-}], \quad (15)$$

Substituting Eqs. (8), (13) and (14) into Eq. (15) and solving yields a single self-consistent value for the Fermi energy at equilibrium that ensures that both charge neutrality and electron/hole mass action are obeyed. From this, the equilibrium concentrations for all defect types, electrons and holes can be calculated.

Table III summarizes the equilibrium concentrations for intrinsic defects in $Sc_2O_3$ under cathode operating conditions (intermediate $\mu_{Sc}^{Sc_2O_3}$ and $\mu_{O}^{Sc_2O_3}$ for both GGA and HSE defect formation energy values). The values for the intrinsic Fermi level $E_{Fermi,intrinsic}$ and intrinsic carrier concentration $n_i$ for a perfect crystal of $Sc_2O_3$ are given as a point of comparison to the values obtained when defects are present, and were calculated using both the GGA and HSE density of states data and the calculated $E_{gap}$ values of 3.9 eV (GGA) and 5.7 eV (HSE). In both the GGA and HSE cases, the dominant defect types are O interstitials on the 16c Wyckoff site, giving rise to an excess of holes in the valence band. The equilibrium hole concentration is roughly 3 orders of magnitude larger than the intrinsic (thermally generated) carrier concentration for GGA and 4 orders of magnitude larger when HSE defect formation energies are used, indicating that under cathode operating conditions bulk $Sc_2O_3$ should behave as a p-type semiconductor.

To the best of our knowledge, there is no experimental data for electron or hole mobility in $Sc_2O_3$. To understand how well $Sc_2O_3$ will conduct under cathode operating conditions solely from intrinsic defects, we use our calculated hole concentrations at equilibrium and calculate ballpark conductivity values based on the mobility of other oxide materials. A survey of mobility values for ZnO, $SnO_2$, $SiO_2$, $In_2O_3$, $ZrO_2$, $UO_2$, BaO and $TiO_2$ yielded values between $10^{-2}$ to $10^{3}$ $cm^2$/V-s.[56-63] From some extrapolation of the data trends presented for the above list of oxides, the mobility at high T for polycrystalline oxide samples is roughly between $10^{-2}$ to 1 $cm^2$/V-s.

Table II. GGA and HSE $\Delta E_{form}^{\alpha}$ for various defect types under the different Sc and O external conditions developed above. Both the GGA and HSE values have been corrected for finite size effects. In the formation energy columns, the GGA value is listed first and the HSE value second. All formation energies are referenced to the VBM.

| Defect | Charge on defect | Kröger-Vink notation | $\Delta E_{form}^{\alpha}$ (Sc rich) (eV) | $\Delta E_{form}^{\alpha}$ (O rich) (eV) | $\Delta E_{form}^{\alpha}$ (intermediate $\mu_{Sc}^{Sc_2O_3}$ and $\mu_{O}^{Sc_2O_3}$ ) (eV) |
|---|---|---|---|---|---|
| $V_{Sc8b}$ | 0 | $V_{Sc}^{x}$ | 13.311 / 16.681 | 3.406 / 6.625 | 6.209 / 9.429 |
| $V_{Sc8b}$ | -1 | $V_{Sc}^{\prime}$ | 13.394 / 17.055 | 3.488 / 6.999 | 6.292 / 8.832 |
| $V_{Sc8b}$ | -2 | $V_{Sc}^{\prime\prime}$ | 13.584 / 17.513 | 3.678 / 7.457 | 6.483 / 9.089 |
| $V_{Sc8b}$ | -3 | $V_{Sc}^{\prime\prime\prime}$ | 14.056 / 18.297 | 4.150 / 8.241 | 6.957 / 9.465 |
| $V_{Sc24d}$ | 0 | $V_{Sc}^{x}$ | 13.317 / 16.413 | 3.412 / 6.357 | 6.216 / 9.161 |
| $V_{Sc24d}$ | -1 | $V_{Sc}^{\prime}$ | 13.355 / 16.794 | 3.449 / 6.738 | 6.254 / 9.542 |
| $V_{Sc24d}$ | -2 | $V_{Sc}^{\prime\prime}$ | 13.493 / 17.258 | 3.587 / 7.202 | 6.393 / 10.006 |
| $V_{Sc24d}$ | -3 | $V_{Sc}^{\prime\prime\prime}$ | 14.232 / 18.477 | 4.327 / 8.421 | 7.138 / 11.224 |
| $V_{Sc24d}$ | -4 | $V_{Sc}^{\prime\prime\prime\prime}$ | 17.742 / n/a | 7.746 / n/a | 10.550 / n/a |
| $V_{O48e}$ | 0 | $V_{O}^{x}$ | 0.837 / 0.944 | 7.441 / 7.648 | 5.571 / 5.779 |
| $V_{O48e}$ | +1 | $V_{O}^{\bullet}$ | -1.900 / -2.555 | 4.704 / 4.149 | 2.835 / 2.280 |
| $V_{O48e}$ | +2 | $V_{O}^{\bullet\bullet}$ | -4.563 / -6.284 | 2.041 / 0.420 | 0.174 / -1.449 |
| $V_{O48e}$ | +3 | $V_{O}^{\bullet\bullet\bullet}$ | -4.496 / n/a | 2.109 / n/a | 0.239 / n/a |
| $Sc_{i,8a}$ | 0 | $Sc_i^x$ | 4.289 / 5.585 | 14.194 / 15.641 | 11.389 / 12.837 |
| $Sc_{i,8a}$ | +1 | $Sc_i^{\bullet}$ | 0.822 / 0.610 | 10.728 / 10.666 | 7.923 / 7.862 |
| $Sc_{i,8a}$ | +2 | $Sc_i^{\bullet\bullet}$ | -2.296 / -3.648 | 7.609 / 6.408 | 4.806 / 3.604 |
| $Sc_{i,8a}$ | +3 | $Sc_i^{\bullet\bullet\bullet}$ | -5.164 / -7.956 | 4.742 / 2.100 | 1.943 / -0.704 |
| $Sc_{i,16c}$ | 0 | $Sc_i^x$ | 4.670 / 6.049 | 14.576 / 16.105 | 11.778 / 13.302 |
| $Sc_{i,16c}$ | +1 | $Sc_i^{\bullet}$ | 1.012 / 0.816 | 10.917 / 10.872 | 8.120 / 8.068 |
| $Sc_{i,16c}$ | +2 | $Sc_i^{\bullet\bullet}$ | -2.587 / -3.933 | 7.319 / 6.123 | 4.519 / 3.319 |
| $Sc_{i,16c}$ | +3 | $Sc_i^{\bullet\bullet\bullet}$ | -5.626 / -8.287 | 4.280 / 1.769 | 1.476 / -1.036 |
| $Sc_{i,16c}$ | +4 | $Sc_i^{\bullet\bullet\bullet\bullet}$ | -5.666 / n/a | 4.330 / n/a | 1.526 / n/a |
| $O_{i,8a}$ | 0 | $O_i^x$ | 9.229 / 10.675 | 2.625 / 3.971 | 4.495 / 5.840 |
| $O_{i,8a}$ | -1 | $O_i'$ | 9.405 / 11.442 | 2.801 / 4.738 | 4.670 / 6.608 |
| $O_{i,8a}$ | -2 | $O_i''$ | 9.809 / 11.882 | 3.205 / 5.178 | 5.074 / 7.047 |
| $O_{i,8a}$ | -3 | $O_i'''$ | 13.506 / 17.173 | 6.902 / 10.470 | 8.771 / 12.338 |
| $O_{i,16c}$ | +1 | $O_i^{\bullet}$ | 8.732 / n/a | 2.128 / n/a | 3.998 / n/a |
| $O_{i,16c}$ | 0 | $O_i^x$ | 8.598 / 10.516 | 1.994 / 3.812 | 3.864 / 5.681 |
| $O_{i,16c}$ | -1 | $O_i'$ | 8.558 / 10.680 | 1.954 / 3.976 | 3.823 / 5.845 |
| $O_{i,16c}$ | -2 | $O_i''$ | 8.937 / 11.271 | 2.333 / 4.567 | 4.203 / 6.436 |
| $O_{i,16c}$ | -3 | $O_i'''$ | 12.388 / 16.159 | 5.784 / 9.455 | 7.654 / 11.324 |
| $O_{i,16c}$ | -4 | $O_i''''$ | 16.123 / n/a | 9.519 / n/a | 11.389 / n/a |

Table III. Summary of results for solving the defect model self-consistently under cathode operating conditions.

| Physical Quantity | GGA Result | HSE Result |
|---|---|---|
| $E_{Fermi,intrinsic}$ (eV) | 2.09 | 2.95 |
| $n_i$ (cm$^{-3}$) | 1.70x10$^{13}$ | 3.83x10$^{9}$ |
| $E_{Fermi,equilibrium}$ (eV) | 1.36 | 2.10 |
| $p_v$ (cm$^{-3}$) | 2.00x10$^{16}$ | 1.39x10$^{13}$ |
| $n_c$ (cm$^{-3}$) | 1.44x10$^{10}$ | 1.06x10$^{6}$ |
| $[O_{i,16c}^{2-}]$ (cm$^{-3}$) | 9.98x10$^{15}$ | 7.06x10$^{12}$ |
| $[V_{O48e}^{2+}]$ (cm$^{-3}$) | 3.49x10$^{10}$ | 1.36x10$^{11}$ |
| $[O_{i,8a}^{2-}]$ (cm$^{-3}$) | 1.10x10$^{12}$ | 9.60x10$^{9}$ |
| $[V_{Sc8b}^{3-}]$ (cm$^{-3}$) | 7.04x10$^{9}$ | 4.52x10$^{8}$ |
| $[V_{Sc24d}^{3-}]$ (cm$^{-3}$) | 3.67x10$^{9}$ | 55.86 |
| $[Sc_{i,16c}^{3+}]$ (cm$^{-3}$) | 0.08 | 1.25 |
| $[Sc_{i,8a}^{3+}]$ (cm$^{-3}$) | 4.22x10$^{-4}$ | 2.52x10$^{-2}$ |

The highest mobility value of $10^3$ cm$^2$/V-s belonged to single crystal ZnO at 50 K, which has a mobility on the order of highly doped Si. The conductivities have been estimated by using the expression for the conductivity of a semiconductor as found in Ref. 55 with different mobility values for the hole densities calculated from both GGA and HSE. These conductivity values are summarized in Table IV. From Table IV we see that the highest calculated conductivity for $Sc_2O_3$ under cathode operating conditions due to intrinsic defects is 3.20 $\Omega^{-1}cm^{-1}$. By comparison, pure Cu has a conductivity of about $1.22 \times 10^5$ $\Omega^{-1}cm^{-1}$ at 1165 K,[64] which is about 5 orders of magnitude higher than the best possible conductivity value for $Sc_2O_3$ from intrinsic defects. In reality, the conductivity of $Sc_2O_3$ is most likely more than 5 orders of magnitude lower than that of Cu, because the mobility of holes in $Sc_2O_3$ will likely be lower than single crystal ZnO at 50K, and the carrier concentration should be closer to that predicted by HSE than GGA due to the increased robustness of the HSE DFT calculations over GGA at predicting defect formation energies.[30,31,65,66]

Table IV. Conductivities for $Sc_2O_3$ using the calculated equilibrium hole concentrations from both GGA and HSE for different mobilities.

| | GGA: $p_v = 2.00 \times 10^{16}$ $h^+/cm^3$ | | HSE: $p_v = 1.39 \times 10^{13}$ $h^+/cm^3$ | |
|---|---|---|---|---|
| Source of Mobility | μ ($cm^2$/V-s) | σ (1/Ω-cm) | μ ($cm^2$/V-s) | σ (1/Ω-cm) |
| Bulk oxide (low) | $10^{-2}$ | $3.20 \times 10^{-5}$ | $10^{-2}$ | $2.22 \times 10^{-8}$ |
| Bulk oxide (high) | $10^{3}$ | 3.20 | $10^{3}$ | $2.22 \times 10^{-3}$ |

Although $Sc_2O_3$ will not behave as a “good” electronic conductor like Cu under operating conditions, it could conduct well *enough* such that the limiting step for thermionic emission lies at the W/$Sc_2O_3$ interface or at the $Sc_2O_3$ surface, and not conduction of electrons through bulk $Sc_2O_3$. To investigate this, we consider the simple circuit model depicted in Fig. 4. This model consists of a series circuit composed of the W substrate, 200 nm $Sc_2O_3$ film, vacuum region, and electron collection anode. This model is essentially the idealized geometry of the cathode experiments detailed in Ref. 7. The voltage drop across the W substrate is considered negligible since W is metallic and hence a good conductor. The emission current density was approximated from data presented in Ref. 67 at 1200 K and the applied voltage of 15 kV was quoted in Ref. 7.

Using this circuit model combined with applying Ohm’s law in Region II ($J = \sigma E$) and invoking current continuity in steady state operation, we use our conductivity values from Table IV to determine the electric field and voltage drop across the $Sc_2O_3$ film. We ascertain whether the conductivity of $Sc_2O_3$ from intrinsic defects is high *enough* by asserting that the electric field drop across the $Sc_2O_3$ film must be well below the dielectric strength of typical insulating materials (so as to not cause breakdown), and that the voltage drop across the film will be imperceptible relative to the total applied voltage.

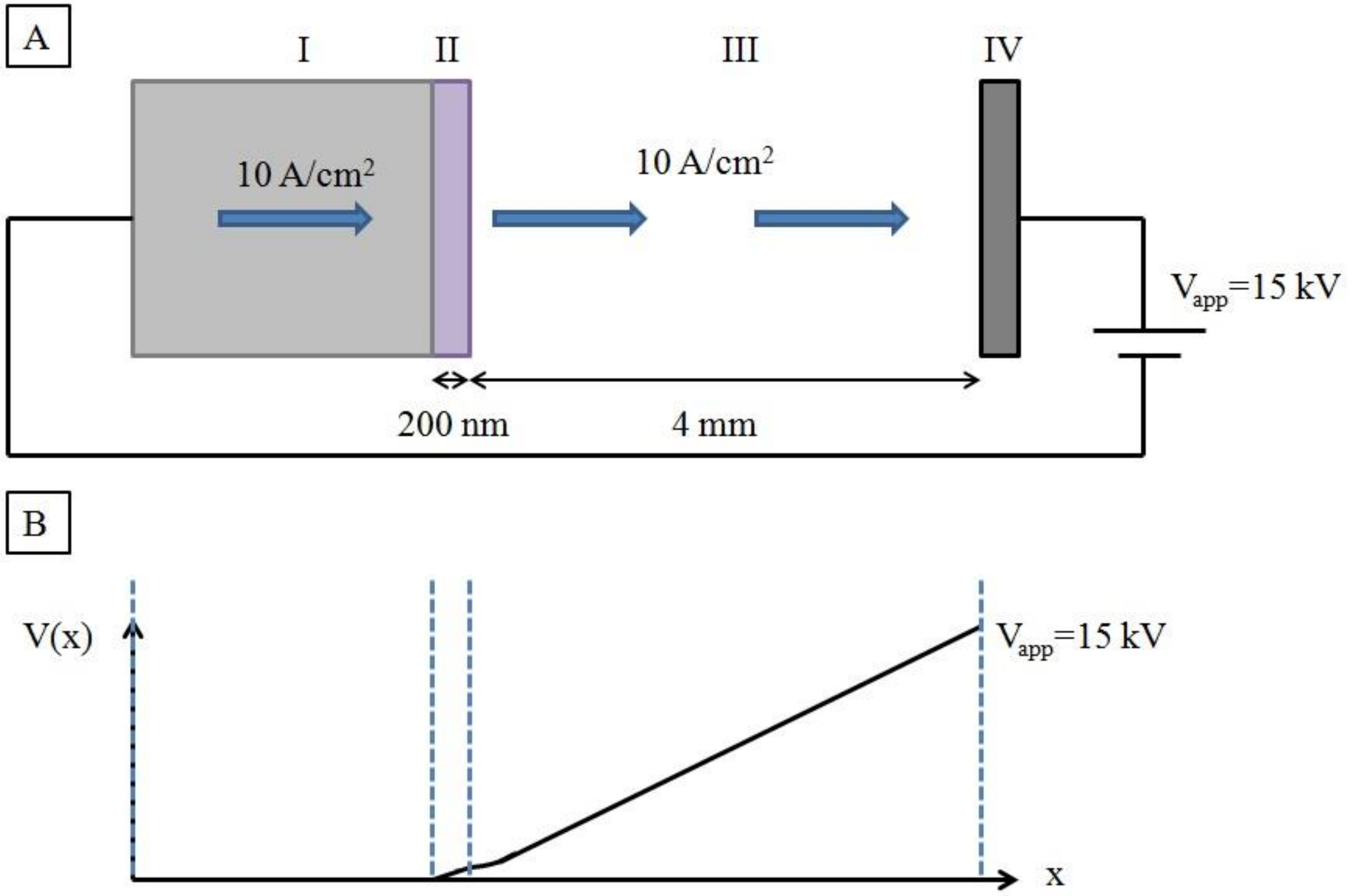

FIG. 4. (Color online) Part A: Setup of scandate cathode series circuit model. Region I is the W substrate, II is the $Sc_2O_3$ thin film, III is the vacuum region, and IV is the anode plate. An approximate current density of 10 A/cm$^2$ flows through both the cathode and vacuum region to the anode. Part B: Approximate potential V(x) at different points in the circuit. It is assumed that the potential drop is linear, except near the cathode emitting surface. Near the surface, space charge depression produces a slight non-linearity in the potential.[68] The dotted lines serve as guides to the eye. Note that dimensions are not to scale.

It is necessary to maintain a small voltage drop across the $Sc_2O_3$ film so that the emitted electrons possess a kinetic energy that is a result of the total applied potential between the cathode and anode. If the voltage drop across the $Sc_2O_3$ film is a large portion of the applied potential, then the cathode will fail to yield significant electron emission. The calculated values of electric field and voltage drop across the film for different calculated conductivities are presented in Table V.

Table V. Calculated electric field and voltage drops across 200 nm $Sc_2O_3$ film based on a series circuit model, calculated conductivities and experimental data.

| GGA: $p_v = 2.00 \times 10^{16}$ $h^+/cm^3$ | | | |
|---|---|---|---|
| μ (cm$^2$/V-s) | σ (1/ Ω-cm) | E (V/cm) | V (Volts) |
| $10^{-2}$ | $3.20 \times 10^{-5}$ | $3.13 \times 10^{5}$ | 6.26 |
| $10^{3}$ | 3.20 | 3.13 | $6.26 \times 10^{-5}$ |
| HSE: $p_v = 1.39 \times 10^{13}$ $h^+/cm^3$ | | | |
| μ (cm$^2$/V-s) | σ (1/ Ω-cm) | E (V/cm) | V (Volts) |
| $10^{-2}$ | $2.22 \times 10^{-8}$ | $4.50 \times 10^{8}$ | $9.00 \times 10^{3}$ |
| $10^{3}$ | $2.22 \times 10^{-3}$ | $4.50 \times 10^{3}$ | 0.09 |

The dielectric strength for insulating materials ranges over one order of magnitude between roughly $5\times10^4$ and $5\times10^5$ V/cm.[69] From this, it is reasonable to assume that electric fields less than $10^5$ V/cm won't cause breakdown of the $Sc_2O_3$ film, which results in a conductivity that must be greater than or equal to $10^{-4}$ $\Omega^{-1}$-cm$^{-1}$ to sustain an emission current density of 10 A/cm$^2$. From Table V, the hole density predicted by HSE yields conductivities that are generally too low to be likely for our system. It is only for the very highest oxide mobilities that a conductivity above $10^{-4}$ $\Omega^{-1}$-cm$^{-1}$ can be obtained. For a typical high-temperature oxide hole mobility of ~1 cm$^2$/V-s the hole density is too small by about a factor of 50. Unless $Sc_2O_3$ has very exceptional hole mobility it is unlikely that the HSE predicted intrinsic carriers provide adequate conductivity for cathode emitter operation. Due to the robustness of the HSE calculation method over GGA at predicting defect formation energies (and hence defect and electron/hole concentrations) it is not reasonable to assume that $Sc_2O_3$ will have a hole concentration as large as $2.00\times10^{16}$ h$^+$/cm$^3$. Rather, we conclude that HSE yields the more accurate value for the hole concentration, and thus from the arguments made above, intrinsic defects are not likely to be solely responsible for the electronic conduction in $Sc_2O_3$ under cathode operating conditions.

Since intrinsic defects likely provide an insufficient number of holes in $Sc_2O_3$ to account for the experimentally observed current densities, we propose that impurity atoms may be responsible for providing the necessary hole density to have sufficiently high conductivity while maintaining physically reasonable values of the mobility, dielectric strength, and voltage drop across the film. Since the $Sc_2O_3$ samples used in scandate cathodes are polycrystalline, other factors besides impurity atoms may also affect the conductivity, such as grain boundaries or regions of off-stoichiometry due to imperfect film growth. Determining the exact nature of these additional factors that affect conduction is beyond the scope of the current study, however we can still estimate the contribution of carriers from a hypothetical impurity atom, with the physically reasonable assumption that each impurity present in the $Sc_2O_3$ will contribute a single charge carrier to increase the conductivity. From our survey of mobility values for different oxide materials, we place a bound on the mobility of $Sc_2O_3$ between $10^{-2}$ to $10^3$ cm$^2$/V-s. Using this bound on mobility in conjunction with the current density, breakdown field, and conductivity requirements developed above, we place a bound on the required hole concentration of $6.25\times10^{16}$ h$^+$/cm$^3$ (for $\mu = 10^{-2}$ cm$^2$/V-s) down to $6.25\times10^{11}$ h$^+$/cm$^3$ (for $\mu = 10^3$ cm$^2$/V-s). With this information, we now estimate the bound on the required impurity concentration needed to raise the $Sc_2O_3$ conductivity to its minimum required value of $10^{-4}$ $\Omega^{-1}$-cm$^{-1}$. $Sc_2O_3$ contains $8.32\times10^{22}$ atoms/cm$^3$, therefore an impurity concentration in the range $7.5\times10^{-6}$ to 0.75 ppm (parts impurity per million lattice sites) is needed to raise the conductivity to $10^{-4}$ $\Omega^{-1}$-cm$^{-1}$ for mobilities between $10^3$ and $10^{-2}$ cm$^2$/V-s. We note that the lowest impurity concentrations in this range are unlikely to be the required concentration as they are consistent with very exceptional mobilities for $Sc_2O_3$ (furthermore, as mentioned above, for such high mobilities the intrinsic defect concentrations are adequate to explain the material's conductivity). If we again consider a typical high-temperature oxide hole mobility of ~1 cm$^2$/V-s an impurity concentration of $7.5\times10^{-3}$ ppm is needed to meet the required conductivity. $Sc_2O_3$ samples used in experimental studies are 99.999% pure at best,[70,71] so it is reasonable to assume that the $Sc_2O_3$ used in thermionic cathodes has, at best, the same purity level. A purity of 99.999% is still impure enough to have a sufficient concentration of impurities present to provide a high enough conductivity for the experimentally observed emission current densities, and this is true for any reasonable mobility of $Sc_2O_3$.

## IV. CONCLUSIONS

The defect energetics for intrinsic point defects in $Sc_2O_3$ have been investigated with DFT-based methods using both the semilocal GGA and hybrid HSE electron exchange and correlation functionals. A defect model was developed whereby the defect, electron, and hole concentrations were calculated by seeking a self-consistent value for the Fermi energy while maintaining overall charge neutrality and electron/hole mass action.

Based on these results, it is expected that $Sc_2O_3$, an insulating material under room temperature and pressure conditions, will behave as a p-type semiconductor under cathode operating conditions of approximately 1200 K and $10^{-10}$ Torr with equilibrium hole densities of $2.00\times10^{16}$ and $1.39\times10^{13}$ $h^+/cm^3$ for GGA and HSE, respectively.

Using these hole densities the conductivity of $Sc_2O_3$ was calculated using a range of mobility data from other oxide materials. A simple series circuit model was used with Ohm's law applied to the oxide film to calculate the conductivity-dependent voltage and electric field drops across a 200 nm film of $Sc_2O_3$. Overall, it was reasoned that the conductivity must be greater than or equal to $10^{-4}$ $\Omega^{-1}$-cm$^{-1}$ to sustain thermionic emission current densities of 10 A/cm$^2$ without a perceivable voltage drop or breakdown of the $Sc_2O_3$ film. Our results indicate that the conductivity of $Sc_2O_3$ solely due to the presence of intrinsic defects in the cathode operating environment is unlikely to be high enough to maintain the magnitude of emitted current densities obtained from experiment. Impurities present in a concentration range of $7.5\times10^{-6}$ – 0.75 ppm are needed to raise the conductivity to $10^{-4}$ $\Omega^{-1}$-cm$^{-1}$ for mobilities between $10^3$ and $10^{-2}$ cm$^2$/V-s, where the impurity concentration of $7.5\times10^{-3}$ ppm (corresponding to a mobility of 1 cm$^2$/V-s) is a reasonable value to consider for typical oxides. Therefore, a very small concentration of impurities can make the conductivity of $Sc_2O_3$ high enough to explain the experimentally observed current densities, and due to the very low value needed such a concentration is easily expected to exist in all experimental $Sc_2O_3$ samples.

## ACKNOWLEDGEMENT

This work was supported by the US Air Force Office of Scientific Research, through grant numbers FA9550-08-0052 and FA9550-11-0299. This work used the Extreme Science and Engineering Discovery Environment (XSEDE), which is supported by National Science Foundation grant number OCI-1053575.

---